\documentclass[twocolumn,showpacs,amsmath,amssymb]{revtex4}
\usepackage{graphicx}
\usepackage{dcolumn}
\usepackage{bm}
\def\br{\begin{eqnarray}}
\def\er{\end{eqnarray}}
\def\be{\begin{equation}}
\def\ee{\end{equation}}

\def\({\left(}
\def\){\right)}

\begin{document}

\title{Light composite Higgs boson from the normalized Bethe-Salpeter equation}
 
\author{A.~Doff$^1$, A.~A.~Natale$^2$ and P.~S.~Rodrigues da Silva$^3$}
\affiliation{$^1$Universidade Tecnol\'ogica Federal do Paran\'a - UTFPR - COMAT
Via do Conhecimento Km 01, 85503-390, Pato Branco - PR, Brazil \\
$^2$Instituto de F\'{\i}sica Te\'orica, UNESP 
Rua Pamplona, 145,
01405-900, S\~ao Paulo - SP,
Brazil \\
$^3$Departamento de F\'{\i}sica, Universidade Federal da
Para\'{\i}ba,
Caixa Postal 5008, 58051-970, Jo\~ao Pessoa - PB, Brazil}

\date{\today}

\begin{abstract}
Scalar composite boson masses have been computed in QCD and Technicolor theories with the help of the homogeneous Bethe-Salpeter equation (BSE),
resulting in a scalar mass that is twice the dynamically generated fermion or technifermion mass ($m_{dyn}$). We show that in the case of
walking (or quasi-conformal) technicolor theories, where the $m_{dyn}$ behavior with the momenta may be quite different from the one predicted
by the standard operator product expansion, this result is incomplete and we must consider the effect of the normalization condition of the
BSE to determine the scalar masses. We compute the composite Higgs boson mass for several groups with technifermions in the fundamental and
higher dimensional representations and comment about the experimental constraints on these theories, which indicate that models based on
walking theories with fermions in the fundamental representation  may, within the limitations of our approach,  have masses quite near the actual direct exclusion limit.
\end{abstract}

\maketitle

\section{Introduction}

The chiral and gauge symmetry breaking in field theories can be promoted by fundamental scalar bosons through the Higgs boson mechanism.
However the main ideas about symmetry breaking and spontaneous generation of fermion and gauge boson masses in field theory were
based on the superconductivity theory. Nambu and Jona-Lasinio proposed one of the first field theoretical models where all the most
important aspects of symmetry breaking and mass generation, as known nowadays, were explored at length \cite{nl}. The model of Ref.\cite{nl}
contains only fermions possessing invariance under chiral symmetry, although this invariance is not respected by the vacuum of the theory
and the fermions acquire a dynamically generated mass ($m_f$). As a consequence of the chiral symmetry breaking by the vacuum the
analysis of the Bethe-Salpeter equation shows the presence of Goldstone bosons. These bosons, when the theory is assumed to be
the effective theory of strongly interacting hadrons, are associated to the pions, which are not true Goldstone bosons when the 
constituent fermions have a small bare mass. Besides these aspects Nambu and Jona-Lasinio also verified that the theory presents
a scalar bound state (the sigma meson) with mass $m_\sigma \approx 2 m_f$.

In Quantum Chromodynamics (QCD) the same mechanism is observed, where the quarks acquire a dynamically generated mass ($m_{dyn}$).
This dynamical mass is usually expected to appear as a solution of the Schwinger-Dyson equation (SDE) for the fermion propagator when the
coupling constant is above a certain critical value. The condition that implies a dynamical mass for quarks breaking the chiral
symmetry is the same one that generates a bound-state massless pion. This happens because the quark propagator SDE is formally the same
equation binding a quark and antiquark into the massless pseudoscalar state at zero momentum transfer (the pion). However, as shown
by Delbourgo and Scadron \cite{ds}, the same similarity of equations happens for the scalar p-wave state of the BSE, indicating the
presence of a bound state with mass $m_\sigma = 2 m_{dyn}$. This scalar meson is the elusive sigma meson \cite{polosa}, that is assumed
to be the Higgs boson of QCD. The basic equations describing such mechanism are: 
\be
\Sigma (p^2) \approx  \Phi_{BS}^P (p,q)|_{q \rightarrow 0} \approx \Phi_{BS}^S (p,q)|_{q^2 = 4 m_{dyn}^2 }\,\,\, ,
\label{eq1}
\ee
where the solution of the fermionic Schwinger-Dyson equation ($\Sigma (p^2)$), that indicates the
generation of a dynamical quark mass and chiral symmetry breaking of QCD, is a solution of 
the homogeneous Bethe-Salpeter equation for a massless pseudoscalar bound state ($\Phi_{BS}^P (p,q)|_{q \rightarrow 0}$),
indicating the existence of Goldstone bosons (pions), and is also a solution of  the 
homogeneous BSE of a scalar p-wave bound state ($\Phi_{BS}^S (p,q)|_{q^2 = 4 m_{dyn}^2 }$), which
implies the existence of the scalar (sigma) boson with the mass described above.  

Non-Abelian gauge theories, if they do not contain fundamental scalar bosons, may undergo the process that we discussed above, where
the scalar boson plays the role of the Higgs boson. In particular, the gauge symmetry breaking of the standard model can also be promoted
through this dynamical mechanism if there are new fermions interacting strongly at the Fermi scale, and models along this idea were named technicolor (TC) models \cite{tc}. It is worth to remember that even if we do not have fermions that condense at the Fermi scale,
the standard model has its gauge symmetry dynamically broken together with the QCD chiral breaking, although do not generate a phenomenologically
viable theory \cite{qs}. 

The same calculation of Ref.\cite{ds} was done by Elias and Scadron in the TC case \cite{es} obtaining a composite Higgs
boson mass given by:
\be
M_H \approx 2 m_{dyn}^{TC} \,\,\, ,
\label{eq2} 
\ee
where $m_{dyn}^{TC}$ is the dynamically generated techniquark mass.
This calculation is simple and elegant, however it was performed before the most recent developments of walking gauge theories~\cite{walk} and assumed a standard operator product expansion behavior (OPE) for the techniquark self-energy. For a standard OPE behavior we mean a self-energy
that behaves as $\Sigma (p) \propto < {\bar \psi} \psi >/p^2$ which appears in a non-Abelian gauge theory which has
a fermion condensate $< {\bar \psi} \psi >$ and an ordinary behavior for the running coupling constant. On the other
hand in an extreme walking gauge theory the standard OPE is modified by a large anomalous dimension ($\gamma$) and the self-energy
behaves as $\Sigma (p) \propto \mu \ln^{-\gamma}(p^2/\mu^2)$, where $\mu$ is the theory's characteristic mass scale.
It was not observed in the work of Ref.\cite{ds} that $m_{dyn}^{TC}$ may vary according the dynamics of the theory that forms the scalar bound state, and the result should be written in terms of known standard model quantities and TC theory gauge group and fermion content. Furthermore, the equalities of Eq.(\ref{eq1}) are obtained from the homogeneous BSE whereas the bound state properties are dictated by the full BSE, i.e. the homogeneous BSE plus its normalization condition, and we observed in Ref.\cite{we} that in walking (or quasi-conformal) technicolor theories the normalization condition of the BSE may affect the result of Eq.(\ref{eq2}).

In this work we study the effect of the normalization condition on the determination of scalar boson masses in dynamically broken gauge theories.
We verify that the normalization condition does not modify the value of the scalar boson mass when its wave function has the asymptotic behavior
exactly as predicted by the OPE. Therefore the determination of the QCD sigma meson mass of Ref.\cite{ds} is not 
modified by the normalization condition. However in walking (or quasi-conformal) gauge theories the asymptotic behavior of fermionic self-energies and
the wave function of scalar bound states are dominated by higher order interactions and are characterized by a much harder decrease with the 
momentum, therefore, in this case, the normalization condition of the BSE do constrain the scalar masses. We determine the correction to 
Eq.(\ref{eq2}) for the composite Higgs boson mass in the case of various gauge theories for fermions in the fundamental and higher 
dimensional representations. We determine the scalar boson masses in the walking regime and comment on the experimental constraints for
these theories.

\section{BSE and the normalization condition}

\par Eq.(\ref{eq1}) tell us that always when the Schwinger-Dyson equation for the fermion propagator has a non-trivial
solution, i.e. a dynamical mass is generated, the homogeneous BSE for (pseudo)scalar bound states, which have identical expressions 
in the Hartree-Fock approximation, also have solutions. However the complete determination of the existence (or not) of bound states is obtained from solutions of the renormalized inhomogeneous BSE. Since the inhomogeneous BSE is quite difficult to solve it is usual to look for the homogeneous solutions
associated with a normalization condition. The BSE normalization condition in the case of a non-Abelian gauge theory is given by \cite{lane2}

\br
2\imath q_{\mu}= \imath^2\!\!\int d^4\!p\, Tr\left\{{\cal P}(p,p+  q)\left[\frac{\partial}{\partial q^{\mu}}F(p,q)\right]{\cal P}(p, p+  q) \right\}\nonumber \\
-\imath^2\!\!\int d^4\!pd^4\!k \,Tr\left\{{\cal P}(k,k +  q)\left[\frac{\partial}{\partial q^{\mu}}K'(p,k,q)\right]{\cal P}(p,  p+ q)\right\} \nonumber
\label{eq43}
\er
where
$$
K'(p,k,q)  = \frac{1}{(2\pi)^4}K(p,k,q)   \,\,\, ,
$$
$$
F(p,q) =  \frac{1}{(2\pi)^4}S^{-1}(p+q) S^{-1}(p) \,\,\, ,
$$ 
and ${\cal P}(p, p + q)$ is a solution of the homogeneous BSE and $K(p,k,q)$ is the fermion-antifermion scattering kernel.
 
\par  When the internal momentum  $q_{\mu} \rightarrow 0$ the wave function ${\cal P}(p, p + q)$ can be determined only through
the knowledge of the fermionic propagator:
\be 
 {\cal P}(p) = S(p)\gamma_{5}\frac{\Sigma(p)}{f'_{\pi}}S(p) \,\, ,
\ee 
\noindent where $\Sigma (p)$ is the fermion self-energy and it should be noticed in the above equation that $f'_{\pi}$ describes the (techni)pion decay constant associated to $n_{d}$ 
fermion doublets, which can be related to the decay constant $f_{\pi}$ in the case $n_{d} = 1$ by $f'^2_{\pi} = n_{d}f^2_{\pi}$.

\par  If we identify $\Sigma(p) \equiv m_{dyn} G(p)$ we can write the normalization condition as
\br
&&2i\left(\frac{f'_{\pi}}{m}\right)^2 q_{\mu} = \frac{i^2}{(2\pi)^4}\times \nonumber \\
&& \left[\int d^4\!p\, Tr{\Big \{}S(p)G(p)\gamma_{5}S(p )\left[\frac{\partial}{\partial q^{\mu}}S^{-1}(p + q) S^{-1}(p)\right]\right. \nonumber \\ 
&& \left. S(p)G(p)\gamma_{5}S(p){\Big \}} + \frac{i^2}{(2\pi)^4}\int d^4\!pd^4\!k \,Tr{\Big \{} S(k)\right. \nonumber \\
&& \left.G(k)\gamma_{5}S(k)\left[\frac{\partial}{\partial q^{\mu}}K(p,k,q)\right]S(p)G(p)\gamma_{5}S(p){\Big \}}\right] \nonumber \\
\label{eq3}
\er 
\par  Eq.(\ref{eq3}) is quite complicated but it can be separated into two parts: 
\be 
2i\left(\frac{f'_{\pi}}{m}\right)^2 q_{\mu}  = I_\mu^{0} + I_\mu^{K} \,\,\, ,
\label{eq30}
\ee
corresponding, respectively, to the two integrals in the right hand side of Eq.(\ref{eq3}).

Eq.(\ref{eq30}) can be further developed, first, if we use the fermion propagator given by
$S(p) = {1}/[{\not{\!\!p} - mG(p)}]$ to write 
\be 
\frac{\partial}{\partial q^{\mu}}S^{-1}(p + q) =  \gamma_{\mu} - m \frac{\partial}{\partial q^{\mu}}G(p + q) \,\,\, ,
\ee
and, secondly, if we consider a specific expression for the fermionic self-energy $(G(p))$. Since we want to make
our calculation the most general as possible we will work with the following expression for $G(p)$ \cite{doff1,doff2,we1}
\be 
G(p) = \left(\frac{m^2}{p^2}\right)^{\alpha}\left[ 1 + bg^2\ln\left(\frac{p^2}{m^2}\right)\right]^{-\gamma(\alpha)} \,\, .
\label{eq4}
\ee
In the above expression we assumed $m = m_{dyn}$, and that $m$ is of the order of $\Lambda$, which is the characteristic QCD(TC) mass scale $\Lambda_{QCD}$($\Lambda_{TC}$),
where the theory breaks the chiral symmetry and form the
composite scalar boson. In Eq.(\ref{eq4}) $g$ is the QCD(TC) running coupling constant, $b$ is the coefficient of $g^3$ term in the renormalization group $\beta (g)$ function, $\gamma (\alpha) = \gamma \cos{(\alpha \pi)}$, where 
\[
\gamma= \frac{3c}{16\pi^2 b}  \,\,\, ,
\] 
and  $c$ is the quadratic Casimir operator given by 
\[
c = \frac{1}{2}\left[C_{2}(R_{1}) +  C_{2}(R_{2}) - C_{2}(R_{3})\right] \,\,\, ,
\] 
where $C_{2}(R_{i})$,  are the Casimir operators for fermions in the representations  $R_{1}$ and 
$R_{2}$ that form a composite boson in the representation $R_{3}$. 
The only restriction on this ansatz is $\gamma > 1/2$ \cite{lane2}, which will be recovered in this work and indicates a condition on the composite wave function normalization. Notice that a standard OPE behavior for $\Sigma (p^2)$ is 
obtained when $\alpha \rightarrow 1$, whereas the extreme walking technicolor solution is obtained when $\alpha \rightarrow 0$.  Note that we performed all the calculations keeping the factor $\alpha$ in Eq.(7). It is at the end of each result that we  take the limit $\alpha = 0$ or $\alpha = 1$, and we can check numerically that the results for intermediate $\alpha$ values are between the ones that we discuss. It is easy to see that for $\alpha =1$ Eq.(7) gives a self-energy roughly equal to $\Sigma (p) \propto m^3/p^2$, which is the behavior predicted by OPE in an ordinary non-Abelian theory (assuming $< {\bar \psi} \psi > \propto m^3$), whereas when $\alpha =0$ the self-energy turns out to be equal to 
$\Sigma (p) \propto m \ln^{-\gamma}(p^2/m^2)$ which is the hardest  behavior allowed for a dynamical fermion
mass in a walking theory\cite{walk,lane2}.

With the ansatz of Eq.(\ref{eq4}) we have
\br
&&\frac{\partial}{\partial q^{\mu}}S^{-1}(p + q) = \gamma_{\mu} + \nonumber \\ 
&&+ m\frac{G(p + q)}{(p + q)^2}\left[\alpha + \gamma(\alpha) bg^2(p+q)\right](p + q)_{\mu}\,\,\, ,
\er
considering the angle approximation to transform the term $G(p + q)/(p + q)^2$ as
\be 
\frac{G(p+q)}{(p + q)^2} = \frac{G(p)}{(p)^2}\Theta(p - q) +  \frac{G(q)}{(q)^2}\Theta(q - p) \,\,\, ,
\ee 
where $\Theta$ is the  Heaviside  step function and we can finally contract Eq.(\ref{eq30}) with $q^\mu$ and compute it at $q^2=M_H^2$ in order to obtain
\br 
 M_{H}^2  =  4m^2&&\hspace{-0.5cm}{\Big\{}-\frac{4n_{f}N}{16\pi^2}\int d^4\!p\frac{m^2G^4(p){\big[}p^2 + m^2G^2(p){\big]^2}}{(p^2 +  m^2G^2(p))^4} \times \nonumber \\
 && \times \frac{1}{(p)^2}(\alpha + \gamma(\alpha) bg^2(p))\left(\frac{m}{f'_{\pi}}\right)^2  + \nonumber  \\
&& + \,\,I^{K}(q^2 = 4m^2,G(p,k),g^2(p,k)) {\Big \}}. 
\label{eq5}  
\er 

Eq.(\ref{eq5}) was determined for a non-Abelian $SU(N)$ theory with the number of fermions equal to $n_f$.
$I^{K}(q^2 = 4m^2,G(p,k),g^2(p,k))$ comes from the second term in the right hand side of Eq.(\ref{eq3}). Actually the simplest truncation
of the kernel $K(p,k;q)$ is the known rainbow-ladder approximation, where
\be
K^{rs}_{tu}(p,k;q\rightarrow 0) = - g^2 D_{\mu\nu}(k-p) \left( \gamma_\mu \frac{\lambda^a}{2} \right)_{tr} \left(\gamma_\nu \frac{\lambda^a}{2} \right)_{su} \,\, .
\label{eq51}
\ee
In this case $\partial_\mu^q K(p,k;q)\equiv 0$ and the second term of the normalization condition (Eq.(\ref{eq3})) does not contribute.
In order to go beyond the rainbow-ladder approximation we should compute $I^K$, however in such case we would have to consider diagrams 
like the ones discussed in Ref.\cite{beyond}, which imply in the calculation of the derivative of the higher order kernel, followed by a large number of integrations. Fortunately, this last contribution is of ${\cal{O}}(g^2(p^2)/4\pi)$ smaller than the one of $I^0$ and it will be neglected in the following. The uncertainty introduced in this last step is not expected to be large since many of the mass generation characteristics are preserved when we go beyond the rainbow approximation\cite{gc}. 

The ansatz of Eq.(\ref{eq4}), when inserted into Eq.(\ref{eq5}) leads to 
\br
M_{H}^2 &=& 4m^2{\Big[}\frac{n_{f}N}{8\pi^2}\left[I_{1}(\alpha)\alpha   
+I_{2}(\alpha)\frac{3cg^2(m)}{16\pi^2}\right]\left(\frac{m}{f'_{\pi}}\right)^2{\Big]}  \,\,\, , \nonumber \\
\er
\noindent where we defined the integrals
\br 
&& I_{1}(\alpha) = \frac{1}{\Gamma(4\gamma)}\int^{\infty}_{0}dz\frac{z^{4\gamma - 1} e^{-z}}{(1 + 4\alpha + \beta z)} \nonumber \\ 
&& I_{2}(\alpha) = \frac{1}{\Gamma(4\gamma + 1)}\int^{\infty}_{0}dz\frac{z^{4\gamma} e^{-z}}{(1 + 4\alpha + \beta z)}, 
\er 
with $\beta = bg^2$.
\par The ratio $\left(\frac{f'_{\pi}}{m}\right)^2$ can now be expressed as 
\be 
\left(\frac{f'_{\pi}}{m}\right)^2 = \frac{n_{f}N}{8\pi^2}\left[ (1 + \frac{\alpha}{2})I_{3}(\alpha) +  \frac{3cg^2(m)}{16\pi^2}I_{4}(\alpha)\right]\,\,\, ,
\ee 
\noindent where
\br
&& I_{3}(\alpha) =  \frac{1}{\Gamma(2\gamma)}\int^{\infty}_{0}dz\frac{z^{2\gamma -  1}e^{-z}}{2\alpha + \beta z} \,\,\, ,\nonumber \\
&& I_{4}(\alpha) = \frac{1}{\Gamma(2\gamma +  1)}\int^{\infty}_{0}dz\frac{z^{2\gamma }e^{-z}}{2\alpha + \beta z}  \,\,\, ,
\er
\noindent so we have
\be 
\frac{n_{f}N}{8\pi^2}(1 + \frac{\alpha}{2})I_{3}(\alpha) = \frac{(1 + \frac{\alpha}{2})}{2Z(\alpha)}.
\ee 
The function $Z(\alpha)$ is the same one that was obtained in Ref.\cite{we1} (see Eq.(38) of that reference) and is equal to
\be
(Z^{(\alpha)})^{-1} \approx  \frac{Nn_{f}}{4\pi^2}\frac{1}{\Gamma(2\gamma)}\int^{\infty}_{0}dz\frac{z^{2\gamma - 1}e^{-z}}{(2\alpha + \beta z)}\,\,  .
\ee

\par The final result for the scalar composite boson mass obtained with the help of the BSE normalization condition can be written as
\br 
M_{H}^2 &=& 4m^2 \left[\frac{I_{1}(\alpha)\alpha + I_{2}(\alpha)\frac{3cg^2(m)}{16\pi^2}}{(1 + \frac{\alpha}{2})I_{3}(\alpha) +  \frac{3cg^2(m)}{16\pi^2}I_{4}(\alpha)}\right] \nonumber \\
&& + {\cal O}(g^2) \,\, ,
\label{eq6}
\er 
where $m$ is the dynamical mass and
the uncertainty of ${\cal O}(g^2)$ in Eq.(\ref{eq6}) is to remember that we neglected the contribution of the term $I^K$. 
As we shall see in the following the term between brackets in Eq.(\ref{eq6}) will reduce the value of the scalar composite mass.

\section{Scalar masses in QCD and TC}

In QCD it is expected that the asymptotic behavior of the dynamical quark self-energy is proportional to $m^3_{dyn}/p^2$ \cite{lane2}.
This means that the self-energy, as well as the wave function of the scalar bound state, decrease very fast with the momentum. This
situation corresponds in our ansatz for $G (p)$ (Eq.(\ref{eq4})) to the case where $\alpha \rightarrow 1$, and Eq.(\ref{eq6}) gives
\be
M_H^{(1)} \equiv m_\sigma = 2 m_{dyn}^{QCD}\big(1+ {\cal{O}}(g^2)    \big) \,\,\, ,
\ee
meaning that the result of Delbourgo and Scadron \cite{ds} is not changed by the BSE normalization condition when the fermionic self-energy
has the asymptotic behavior predicted by the standard OPE. This result is easy to understand because the integrals of the BSE
normalization condition are extremely convergent in this case and no extra condition appears in this situation.

The most interesting case is when $\alpha \rightarrow 0$, which is the case of the extreme walking (or quasi-conformal) technicolor theories.
In this case Eq.(\ref{eq6}) gives
\be 
 M_{H}^{2(0)}  \approx 4 m^2 \left( \frac{1}{4}\frac{bg^2(m)(2\gamma - 1)}{(1 + \frac{bg^2(m)(2\gamma - 1)}{2})}\right).
\ee 
\par Notice that in order to have a positive mass we must have $(2\gamma - 1) > 0$, in such a way that we recover Lane's condition \cite{lane2}, i.e.
\be 
\gamma > \frac{1}{2}.
\ee

At this point it is important to remember that the dynamical mass ($m$) depends on the dynamics of a particular theory and, as
emphasized in Ref.\cite{we,fre}, it should be written in terms of known standard model quantities and TC theory gauge group and
fermion content. The way 
this is accomplished follows the work of Ref.~\cite{we}: $m$ will be related to $F_\Pi$ (the  Technipion decay 
constant), and this last one will be related to the vacuum expectation value (VEV) of the Standard Model through
\be
\frac{g_w^2 n_F F_\Pi^2}{8}  = \frac{g_w^2 v^2}{4} = M_W^2 \,\,\, ,
\label{eq33}
\ee
where $g_w$ is the weak coupling constant, $v \sim 246 GeV$ is the Standard Model VEV and $F_\Pi$ is obtained from the Pagels and Stokar relation~\cite{pagels},  
\be 
F^2_{\Pi} = \frac{N_{{}_{TC}}}{4\pi^2}\int\!\!\frac{dp^2p^2}{(p^2 + \Sigma^2(p^2))^2}\!\!\left[\Sigma^2(p^2) - \frac{p^2}{2}\frac{d\Sigma(p^2)}{dp^2}\Sigma(p^2)\right] \,\,\, .
\label{eq34}
\ee 
Where we are also changing our notation ($n_f \rightarrow n_F$, $N\rightarrow N_{TC}$, $f_\pi \rightarrow F_\Pi$) because in the following we shall refer only to TC theories.

The relation between $F_\Pi$ and $m_{dyn}$ will depend strongly on
the $\Sigma (p^2)$ behavior described by Eq.(\ref{eq2}), and the dynamical masses
in the limits $\alpha =0 $ and $\alpha = 1$ will be given by
\br
&& m^{(0)} \approx  \left[v\left(\frac{8 \pi^2  bg^2(2\gamma -1)}{N_{TC}n_F} \right)^{1/2}\right]
\label{eq410} \\
&& m^{(1)} \approx \left[\sqrt{\frac{4}{3}}v\left( \frac{8\pi^2}{N_{TC}n_{F}}\right)^{1/2}\right].
\label{eq41}
\er 
Therefore, with the help of Eq.(5) of \cite{we} we end up with the following expression for the scalar
composite mass:
\br
M_{H}^{2(0)}  &\approx & 4 v^2\left(\frac{8 \pi^2  bg^2(m)(2\gamma -1)}{N_{TC}n_F} \right) \nonumber \\
&& \times \left( \frac{1}{4}\frac{bg^2(m)(2\gamma - 1)}{(1 + \frac{bg^2(m)(2\gamma - 1)}{2})}\right),
\label{eqf}
\er
where $v \sim 246 GeV$ is the Standard Model vacuum expectation value and we are considering a $SU(N_{TC})$ group with $n_F$ technifermions.

\par The result of Eq.(\ref{eqf}) should be compared to the one of Ref.\cite{we1}, that was obtained from an effective action for composite operators,
and which, if we neglect the contribution from the top quark, gives the following result 
\be 
m^{2}_{H} \approx 2\lambda^{(0)}_{4VR}\left(\frac{\lambda^{(0)}_{4VR}}{\lambda^{(0)}_{6VR}}\right) \,\,\, , 
\label{eq7}
\ee 
\noindent where the  $\lambda's$ couplings appearing in Eq.(\ref{eq7}) are equal to:
\br 
&&\lambda^{(0)}_{4VR} = \frac{N_{TC} n_{F}}{4\pi^2}[Z^{(0)}]^2 \left(\frac{1}{bg^2(4\gamma - 1)} +\frac{1}{2}\right)\, , \nonumber \\
&& \lambda^{(0)}_{6VR} = \frac{N_{TC} n_{F}}{4\pi^2}\frac{[Z^{(0)}]^3}{m^2}  \, , 
\er
\noindent and
\be 
Z^{(0)} = \frac{4 \pi^2 bg^2(2\gamma -1)}{N_{TC}n_F}.
\label{eq71}
\ee 
\par
\begin{figure}[h]
\centering
\includegraphics[width=0.7\columnwidth]{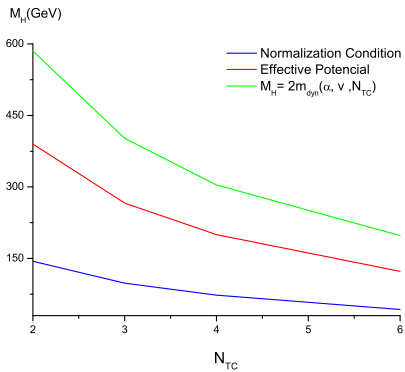}
\caption[dummy0]{ This figure shows the scalar masses computed for extreme  walking $SU(N_{TC})$ gauge theories in the case where  the fermions  are in the fundamental representation. The curves depicted in green (upper),  red (middle) and blue (lower) were computed using respectively  Eqs.(\ref{eq2}),(\ref{eq7}),(\ref{eqf}), and we used the following  number of fermions : $n_F = 8,11,14,18$ for $SU(2)$ to
$SU(6)$ respectively.} 
\label{figmass}
\end{figure}

 In the calculation of the effective action \cite{we1} it was assumed that $c\alpha_{TC} \approx \frac{\pi}{3}$, and the same will be done here.
This choice is the same one proposed in Ref.\cite{sk}, seems to be in agreement with recent estimates of the infrared coupling in the QCD case\cite{Cornwall} and is necessary in order to compare the present calculation with the one of Ref.\cite{we1}.
The dynamical mass can be related with the vacuum expectation value of the Higgs field in the Standard Model through the relation
\be 
v^2 = \frac{n_{F}}{2}F^2_{\Pi} = (1 + \frac{\alpha}{2})\frac{m^2}{2Z(\alpha)}, 
\ee 
\noindent and the technipion decay constant ($F_\Pi$) was obtained from the Pagels and Stokar relation \cite{pagels},
as described in Ref.\cite{we}. $Z(\alpha)$ is a factor computed in Ref.\cite{we1}, which in the limit $\alpha \rightarrow 0$ is reduced
exactly to Eq.(\ref{eq71}). We can now present in Fig.(\ref{figmass}) the masses that we have obtained in the limit $\alpha \rightarrow 0$, that is the limit of walking (or quasi-conformal) theories.

The curves of Fig.(\ref{figmass}) were obtained for extreme walking (or quasi-conformal) $SU(N_{TC})$ gauge theories in the case where
the fermions are in the fundamental representation. These curves  represent three different calculations of the scalar boson masses.
The curve depicted in green correspond to the result obtained with  Eq.(\ref{eq2}), the curve in red was obtained from the effective
action calculation of Ref.\cite{we} and show the masses computed with the help of Eq.(\ref{eq7}), the values of masses indicated
in blue are the ones computed in this work and given by Eq.(\ref{eqf}). We credit the larger mass values obtained in Ref.\cite{we1}
to the fact that the effective action is able to capture some of the non-linear effects present in the SDE (or BSE) and due to the
process of normalization of the effective composite field.

\section{Higgs mass with fermions in higher dimensional representations}

Equation (\ref{eqf}) is a quite general expression to be used in the case of extreme walking (or quasi-conformal) gauge theories, where
the factor $N_{TC}$ has to be changed by $C_2(G)$ when dealing with groups other than $SU(N_{TC})$ technicolor. Furthermore we
have dealt only with gauge theories where the
fermions are in the fundamental representation, but we can use Eq.(\ref{eqf}) to obtain scalar masses when the fermions
belong to higher dimensional representations, as well as for unitary groups other than $SU(N_{TC})$. In order to do so we just have to
compute the coefficients $b$, $\gamma$ for the appropriate groups and number of technifermions ($n_F$). The advantage of working with
higher dimensional fermionic representations has been 
extensively advocated by Sannino and collaborators \cite{san,san2}, and the most
important one is that we can obtain walking, or quasi-conformal, TC theories with a small number of technifermions and in conformity with
high precision standard model data.

The number of fermions in different group representations that lead to a walking (or quasi-conformal) gauge theory can be obtained
looking at the zero of the two-loop $\beta (g^2)$ function, which is given by
\be
\beta (g) = -\beta_0 \frac{g^3}{(4\pi)^2} - \beta_1 \frac{g^5}{(4\pi)^4} \,\,\, ,
\ee
where
\be
\beta_0 = (4\pi)^2 b= \frac{11}{3} C_2(G) - \frac{4}{3}T(R)n_F (R) \,\,\, , 
\ee
and 
\be
\beta_1 = \left[\frac{34}{3}C_2^2(G)-\frac{20}{3}C_2(G)T(R)n_F-4C_2(R)T(R)n_F \right] \,\,\, .
\ee 
We will consider only fermions that are in the representation $R$ and condensate into a singlet state, for which the constant $c$ that enters in the
definition of the parameter $\gamma$ of Eq.(7) is given by
\be
c=C_2(R),
\ee
remembering that $C_2(R)I=T^a_RT^a_R$ and $C_2(R)d(R)=T(R)d(G)$,where $d(R)$ is the dimension of the representation $R$, while the label $G$ refers to the adjoint representation.

Notice that the walking (or quasi-conformal) theory is not necessarily specified by the zero of the $\beta (g)$ function. Actually
there is a window of number of flavors that characterize a quasi-conformal (or walking) theory. This point has been discussed by Sannino
and collaborators\cite{san2}, and they verified that the number of fermions below which the theory undergoes chiral symmetry breaking while quasi-conformal and asymptotically free can be obtained from the following expression:
\be
n_F = \frac{17C_2(G)+66C_2(R)}{10C_2((G)+30C_2(R)}\frac{C_2(G)}{T(R)}\,.
\label{NFII}
\ee
\begin{table}[t]
\begin{ruledtabular}
\begin{tabular}{cccc}
 Group& Representation& $n_F$& Higgs mass (GeV) \\ \\
\hline
   SU(2) &   F &     $8$  & 142   \\ 
   SU(3) &   F &     $12$ & 106   \\ 
   SU(2) &   G &     $2$  & 414   \\ 
   SU(3) &   G &     $2$  & 338   \\ 
   SU(4) &   G &     $2$  & 293   \\ 
   SU(5) &   G &     $2$  & 262   \\ 
   SU(6) &   G &     $2$  & 239   \\ 
   SU(2) &  $S_2$ &  $2$  & 414   \\
   SU(3) &  $S_2$ &  $2$  & 320   \\
   SU(4) &  $S_2$ &  $2$  & 267   \\
   SU(5) &  $S_2$ &  $2$  & 233   \\
   SU(6) &  $S_2$ &  $4$  & 197   \\
   SU(7) &  $S_2$ &  $4$  & 179   \\
   SU(3) &  $A_2$ &  $12$ & 106   \\                      
   SU(4) &  $A_2$ &  $8$  & 130   \\                      
   SU(5) &  $A_2$ &  $6$  & 130   \\                      
   SU(6) &  $A_2$ &  $6$  & 131   \\                      
   SU(7) &  $A_2$ &  $6$  & 128   \\                      
   Sp(2) &  $F$   &  $8$  & 142   \\
   Sp(4) &  $F$   &  $12$ & 102   \\
   Sp(2) &  $G$   &  $2$  & 414   \\
   Sp(4) &  $G$   &  $2$  & 338   \\
   Sp(6) &  $G$   &  $2$  & 293   \\
   Sp(8) &  $G$   &  $2$  & 262   \\
   Sp(10) &  $G = S_2$   &  $2$  & 239  \\
   Sp(12) &  $G = S_2$   &  $2$  & 221   \\
   Sp(14) &  $G = S_2$   &  $2$  & 207   \\
   Sp(4) &  $A_2$   &  $6$  & 177   \\
   Sp(6) &  $A_2$   &  $4$  & 194   \\
   Sp(8) &  $A_2$   &  $4$  & 199   \\
   Sp(10) &  $A_2$   &  $3$  & 188   \\
   Sp(12) &  $A_2$   &  $3$  & 183   \\
   Sp(14) &  $A_2$   &  $3$  & 176   \\
   SO(6)  &  $F$     &  $8$  & 130   \\
   SO(7)  &  $F$     &  $10$  & 103   \\
   SO(6)  &  $G = A_2$     &  $2$  & 293   \\
   SO(7)  &  $G = A_2$     &  $2$  & 262   \\
   SO(8)  &  $G = A_2$     &  $2$  & 239   \\
   SO(9)  &  $G = A_2$     &  $2$  & 221   \\
   SO(10)  &  $G = A_2$     &  $2$  & 207   \\
   SO(14)  &  $S_2$     &  $2$  & 197   \\
   SO(15)  &  $S_2$     &  $2$  & 187   \\
   SO(16)  &  $S_2$     &  $2$  & 179   \\
   SO(17)  &  $S_2$     &  $2$  & 172   \\
\end{tabular}
\end{ruledtabular}
\vspace*{-0.1cm}
\label{atab3}
\caption{ Higgs mass, $M_H$, and number of Dirac fermions, $n_F$, used to compute the Higgs mass in some representantions of $SU(N)$, $Sp(2N)$ and $SO(N)$. The number of fermions considered are such that they are on the border of the conformal window for the representations $F$ (fundamental),
$G$ (adjoint), $S_2$ (2-index symmetric) and $A_2$ (2-index antisymmetric). For the $SO(N)$ we considered only those $N_{TC}$ for wich we could have at least 2 Dirac fermions on the border of the conformal window.}
\end{table}     

\begin{figure}[h]
\hspace*{-1.0cm}
\centering
\includegraphics[width=1.1\columnwidth]{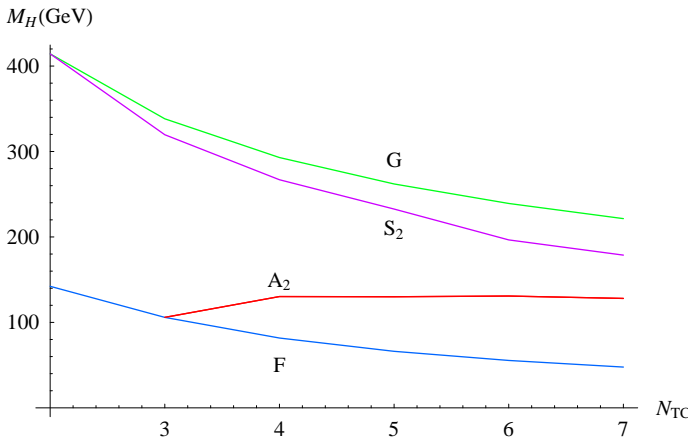}
\caption[dummy0]{ Higgs mass for $SU(N_{TC})$ in the representations $F$ (fundamental),
$G$ (adjoint), $S_2$ (2-index symmetric) and $A_2$ (2-index antisymmetric).}
\label{fig:H1}
\end{figure}
\begin{figure}[h]
\hspace*{-1.0cm}
\centering
\includegraphics[width=1.1\columnwidth]{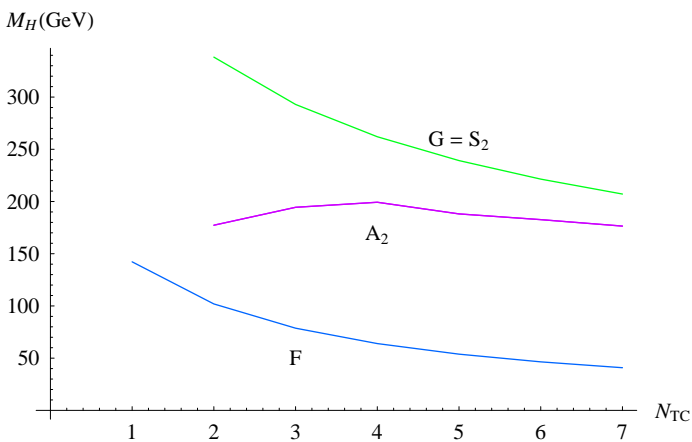}
\caption[dummy0]{Higgs mass for $Sp(2N_{TC})$ in the representations $F$ (fundamental),
$G$ (adjoint), $S_2$ (2-index symmetric) and $A_2$ (2-index antisymmetric).}
\label{fig:H2}
\end{figure}
\begin{figure}[h]
\hspace*{-1.0cm}
\centering
\includegraphics[width=1.1\columnwidth]{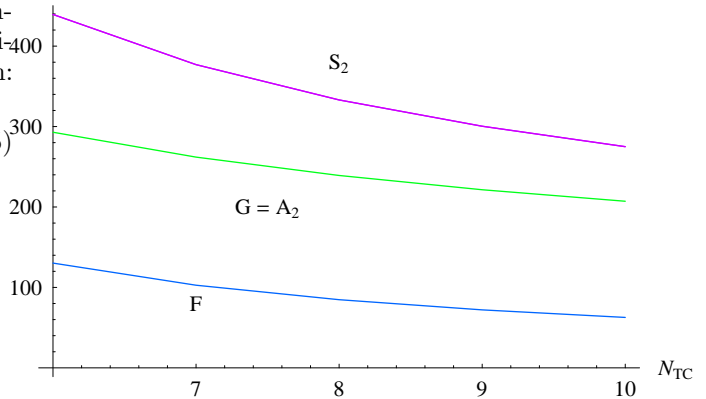}
\caption[dummy0]{Higgs mass for $SO(N_{TC})$ in the representations $F$ (fundamental),
$G$ (adjoint), $S_2$ (2-index symmetric) and $A_2$ (2-index antisymmetric).}
\label{fig:H3}
\end{figure}

Considering the two-loop beta function and using the number of fermion on the border of the conformal window, Eq.(\ref{NFII}) , in order to still have chiral symmetry breaking  leading to the extreme case of walking theories, we obtained the Higgs masses when technifermions are in higher dimensional representations as shown in Table I. We also show the behavior of Higgs mass as a function of $N_{TC}$ in some representations of $SU(N_{TC})$, $Sp(2N_{TC})$ and $SO(N_{TC})$ in Figures (\ref{fig:H1}),(\ref{fig:H2}), and (\ref{fig:H3}).
  
The mass values shown in Table I and in the Figures (\ref{fig:H1}),(\ref{fig:H2}), and (\ref{fig:H3}) are quite light when compared with the
ones determined in the early work of Ref.\cite{es}. These values may still have contributions from the larger order corrections to
Eq.(\ref{eq5}), which are very difficult to calculate, but based on the work of Ref.\cite{beyond} we can at least say that the
sign of the contribution is positive in the sense of increasing the scalar masses. The scalar boson
will interact with the ordinary fermions, and, in particular, strongly with the top quark; this interaction will give a new contribution to the
scalar mass that has been computed in Ref.\cite{we1}, which will also increase all the scalar masses that we have calculated by a factor of the same
order, depending more on the top-scalar coupling than anything else. Other sources of contributions to the scalar masses are more model
dependent such as the fact that a composite scalar boson could mix with other scalars, formed, for instance,
by technigluons, which is a problem already discussed for the sigma meson in QCD~\cite{polosa}, but not taken into account here. Techniquarks may have a current mass and will also introduce an extra contribution to the SDE solution and modify our prediction. Nevertheless all these are probably minor effects that should not change drastically the mass values that we have shown here.

There are experimental constraints on the models shown in Table I. Since some of the obtained masses are very light we have a strong
constraint coming from the lower direct bound on the Higgs boson mass \cite{pdg}:
\[
M_H \geq 114.4 \,\,\, GeV \,\, ,
\]
and a small window of masses above this value has also been constrained by Fermilab data \cite{cdf}. These values when compared to the
ones in Table I and also Figures (\ref{fig:H1}),(\ref{fig:H2}), and (\ref{fig:H3}) show how difficult is to build a realistic model for the dynamical symmetry breaking of the standard model with walking (or quasi-conformal) gauge theories when the fermions are in the fundamental representation. No matter the walking technicolor gauge group is $SU(N_{TC})$, $Sp(2N_{TC})$ or $SO(N_{TC})$ with fermions in the fundamental representation the scalar mass turns out to be very small and, if not excluded, it is on the verge of being tested by future experiments. 

Looking at Figures (\ref{fig:H1}),(\ref{fig:H2}), and (\ref{fig:H3}) we see that when
the fermions are in higher dimensional representations the $SU(N)$,
$Sp(2N)$ or $SO(N)$ theories in general lead to scalar masses above the present experimental limit. This is certainly the case for fermions in
the adjoint and 2-index symmetric representation for all these theories. It is interesting that in the case of technifermions in the
2-index antisymmetric representation of $SU(N_{TC})$ and $Sp(2N_{TC})$
the scalar masses have an almost stable value as we increase $N_{TC}$ while we maintain the walking behavior.

A strong experimental constraint for walking gauge theories with fermions in the fundamental representation also comes from
the limits on the $S$ parameter \cite{peskin}, whose
perturbative expression is
\be
S=\frac{1}{6\pi}\frac{n_F}{2}d(R) \,\, ,
\label{s1}
\ee
where $d(R)$ is the dimension of the representation $R$, obtained from
\[
d(R)=T(R)d(G)/C_2(R) \,\, .
\]
For fermions in the fundamental representation the walking (or quasi-conformal)
behavior is obtained only with a large number of fermions, but this imply large $S$ values.
As discussed by Dietrich and Sannino \cite{san2}, we can impose arbitrarily the limit $S<\pi^{-1}$ on the models
of Table I in order to be in accordance with the experimental limits \cite{pdg}. These constraints severely
limit the possibility of viable models of walking technicolor gauge theories with fermions in the fundamental representation.
We still have possible models with fermions in higher dimensional representations in order to discuss a viable standard model
dynamical symmetry breaking. For instance, in the $SU(N_{TC})$ case, for fermions in the adjoint and 2-index symmetric and
antisymmetric representations, the models that survive the constraint imposed by the $S$ value are the same ones described
by Dietrich and Sannino \cite{san2}. We have not considered the possibility of partial electroweakly gauged technicolor, with or without 
mixed representations. This case deserves  further study and most  certainly may  modify the mass values that we obtained.

\section{Conclusions}

The scalar boson masses generated in dynamically broken gauge theories have been computed in technicolor theories with the help of
the Bethe-Salpeter equation. We have shown that these masses must be written in terms of measurable standard model quantities and
group theoretical factors, and, particularly in the case of walking (or quasi-conformal) technicolor gauge theories, we verified that the mass calculation in the BSE approach must be supplemented with the BSE normalization condition. We recovered also a constraint on the asymptotic behavior
of the scalar wave function obtained many years ago by Lane \cite{lane2}.

We obtained an expression for the scalar composite mass valid for any gauge group and fermion representation. The scalar masses
that appear in the walking limit are much lighter than the old estimate of Ref.\cite{es}. Considering the direct limit on the
Higgs boson mass and the bound determined by the high precision standard model data it is possible to see that a technicolor model with fermions in the fundamental representation, within our approximations, will have a quite light Higgs mass,  which is on the verge of the actual direct exclusion limit and may be soon assessed by the LHC. As can be seen in Figures (\ref{fig:H1}),(\ref{fig:H2}), and (\ref{fig:H3}), when the technifermions are in higher dimensional representations (adjoint, 2-index symmetric and antisymmetric) the scalar masses turn out to be larger, and the models that survive the limit imposed by the experimental value of the $S$ parameter are the same ones already discussed in the literature. It would be interesting to apply the formalism that we discussed
here to the case of partial electroweakly gauged technicolor, with or without mixed representations, from which we may expect different
spectra of masses.

It is quite interesting that the scalar composite masses can be computed under certain controllable approximations, as in the Bethe-Salpeter approach, and the results shown here confirm the ones obtained in a more complicated calculation as the one of the effective action of Ref.\cite{we1}, although there are still differences when comparing the methods, which can be considered natural considering the complexity of these theories. We finally stress the importance of considering the BSE normalization condition when computing scalar masses in the case of walking gauge theories through the BSE approach.

\section*{Acknowledgments}

We thank A. C. Aguilar for useful discussions. This research was partially supported by the Conselho Nacional de Desenvolvimento Cient\'{\i}fico e Tecnol\'ogico (CNPq).


\begin {thebibliography}{99}

\bibitem{nl} Y. Nambu and G. Jona-Lasinio, {\it Phys. Rev.} {\bf 122}, 345  (1961).
\bibitem{ds} R. Delbourgo and M. D. Scadron, {\it Phys. Rev. Lett.} {\bf 48}, 379 (1982).
\bibitem{polosa} N. A. Tornqvist and M. Roos, {\it Phys. Rev. Lett.} {\bf 76}, 1575 (1996);
N. A. Tornqvist and A. D. Polosa, {\it Nucl.Phys. A} {\bf 692}, 259 (2001); {\it Frascati Phys.Ser.} {\bf 20}, 385 (2000);
A. D. Polosa, N. A. Tornqvist, M. D. Scadron and V. Elias, {\it Mod. Phys. Lett. A} {\bf 17}, 569 (2002). 
\bibitem{tc} S. Weinberg, {\it Phys. Rev. D} {\bf 19}, 1277 (1979); L. Susskind, {\it Phys. Rev. D} {\bf 20}, 2619 (1979).
\bibitem{qs} C. Quigg and R. Shrock, hep-ph/0901.3958.
\bibitem{es} V. Elias and M. D. Scadron, {\it Phys. Rev. Lett.} {\bf 53}, 1129 (1984).
\bibitem{walk} B. Holdom, {\it Phys. Rev.} {\bf D24},1441 (1981);{\it Phys. Lett.}
{\bf B150}, 301 (1985); T. Appelquist, D. Karabali and L. C. R.
Wijewardhana, {\it Phys. Rev. Lett.} {\bf 57}, 957 (1986); T. Appelquist and
L. C. R. Wijewardhana, {\it Phys. Rev.} {\bf D36}, 568 (1987); K. Yamawaki, M.
Bando and K.I. Matumoto, {\it Phys. Rev. Lett.} {\bf 56}, 1335 (1986); T. Akiba
and T. Yanagida, {\it Phys. Lett.} {\bf B169}, 432 (1986).
\bibitem{we} A. Doff and A. A. Natale,  hep-ph/0902.2379 (to appear in Phys.Lett.B).
\bibitem{lane2} K. Lane, {\it Phys. Rev.} {\bf D10}, 2605 (1974).
\bibitem{fre} J.-M. Fr\`ere, {\it Phys. Rev. D} {\bf 35}, 2625 (1987).
\bibitem{doff1}A. Doff and A. A. Natale, {\it Phys. Lett. } {\bf B537}, 275 (2002).
\bibitem{doff2} A. Doff and A. A. Natale, {\it Phys. Rev.} {\bf D68}, 077702 (2003).
\bibitem{we1} A. Doff, A. A. Natale and P. S. Rodrigues da Silva, {\it Phys. Rev. D} {\bf 77}, 075012 (2008).
\bibitem{sk} S. Raby, S. Dimopoulos and L. Susskind, Nucl. Phys. B {\bf 169}, 373 (1980); A. A. Natale, Nucl. Phys. B {\bf 226}, 365 (1983).
\bibitem{Cornwall} J. M. Cornwall, hep-ph/0812.0359. 
\bibitem{pagels} H. Pagels and S. Stokar, {\it Phys. Rev. D} {\bf 20}, 2947 (1979).
\bibitem{beyond} P. Watson, W. Cassing and P. C. Tandy, {\it Few Body Syst.} {\bf 35}, 129 (2004); H. H. Matevosyan,
A. W. Thomas and P. C. Tandy, {\it Phys. Rev.} {\bf C75}, 045201 (2007); C. S. Fischer and R. Williams {\it Phys. Rev.} {\bf D78}, 074006 (2008).
\bibitem{gc} A. G. Cohen and H. Georgi, Nucl. Phys. B {\bf 314}, 7 (1989).
\bibitem{san} F. Sannino, {\it Int. J. Mod. Phys. A} {\bf 20}, 6133 (2005); D. D. Dietrich, F. Sannino and K. Tuominen, {\it Phys. Rev. D} {\bf 72}, 055001 (2005);  N. Evans and F. Sannino, hep-ph/0512080; D. D. Dietrich, F. Sannino and K. Tuominen, {\it Phys. Rev. D} {\bf 73}, 037701 (2006);  R. Foadi, M. T. Frandsen, T. A. Ryttov and F. Sannino, {\it Phys. Rev. D} {\bf 76}, 055005 (2007); R. Foadi, M. T. Frandsen and F. Sannino, hep-ph/0712.1948.
\bibitem{san2} D. D. Dietrich and F. Sannino, {\it Phys. Rev. D} {\bf 75}, 085018 (2007); T. A. Ryttov and F. Sannino, {\it Phys. Rev. D} {\bf 76}, 105004 (2007); F. Sannino, hep-ph/0902.3494.
\bibitem{pdg} W.-M. Yao et al. (Particle Data Group), {\it J. Phys. G} {\bf 33}, 1 (2006).
\bibitem{cdf} $http://www-cdf.fnal.gov/\-physics/\-new/\-hdg/\-results/\-combcdf_090116/$
\bibitem{peskin} M. E. Peskin and T. Takeuchi, {\it Phys. Rev. Lett.} {\bf 65}, 964 (1990).

\end {thebibliography}
\end{document}